\documentclass[aps,prd,showpacs,preprintnumbers]{revtex4}%twocolumn
\usepackage{graphicx}
\usepackage{epsfig}
\usepackage{amsmath}
\usepackage{subfigure}
\usepackage{ulem}
\usepackage{color}
\usepackage{latexsym} % Gets \Box etc
\usepackage{amssymb}  % \gtrsim, \geqslant, etc etc: see amsguide.ps
\usepackage{amsfonts} % \mathfrak and \mathbb{x} (Blackboard bold)
\usepackage{amsbsy}   % \pmb and \boldsymbol
\usepackage{multirow} % for multirows in tables
\usepackage{pstricks}
\usepackage{slashed}
\usepackage{dcolumn}
\usepackage{xparse}
\usepackage{bm}
\newcommand{\al}{\alpha}
\newcommand{\be}{\beta}
\newcommand{\ga}{\gamma}
\newcommand{\Ga}{\Gamma}

\newcommand{\ep}{\varepsilon}
\newcommand{\eps}{\epsilon}

\newcommand{\la}{\lambda}
\newcommand{\La}{\Lambda}

\newcommand{\si}{\sigma}

\newcommand{\om}{\omega}

\newcommand{\Th}{\Theta}
\newcommand{\beq}{\begin{equation}}
\newcommand{\eeq}{\end{equation}}
\newcommand{\ba}{\begin{array}}
\newcommand{\ea}{\end{array}}
\newcommand{\bea}{\begin{eqnarray}}
\newcommand{\eea}{\end{eqnarray}}
\newcommand{\bi}{\begin{itemize}}
\newcommand{\ei}{\end{itemize}}
\newcommand{\ben}{\begin{enumerate}}
\newcommand{\een}{\end{enumerate}}
\newcommand{\bc}{\begin{center}}
\newcommand{\ec}{\end{center}}
\newcommand{\bl}{\begin{flushleft}}
\newcommand{\el}{\end{flushleft}}
\newcommand{\br}{\begin{flushright}}
\newcommand{\er}{\end{flushright}}

\newcommand{\nn}{\nonumber \\}
\newcommand\Eqn[1]{Eq.~(\ref{#1})}  % includes ``Eq.'' in front
\newcommand{\mr}{\mathrm}
\newcommand{\mb}{\mathbf}

\newcommand{\mc}{\mathcal}
\newcommand{\mi}{\mathop{}\!i}
\newcommand{\me}{\mathop{}\!e}
\newcommand{\dif}{\mathop{}\!d}
\newcommand{\p}{\partial}

\newcommand{\Tr}{\hbox{Tr}}

\renewcommand{\Re}{{\mathrm{Re}}\,}
\renewcommand{\Im}{{\mathrm{Im}}\,}

\newcommand{\<}{\langle}
\renewcommand{\>}{\rangle}   % LaTeX: \> already defined
\renewcommand{\l}{\left}
\renewcommand{\r}{\right}

\newcommand\comment[1]{ \hbox{[{\it Comment suppressed here.}\/]} }
\newcommand\hide[1]{}
\newcommand{\skipover}[1]{}
\newcommand{\sign}{\,\mbox{sign}}

\definecolor{red}{rgb}{0.7,0,0}
\definecolor{green}{rgb}{0,0.5,0}

\newcommand{\kp}{k_{\shortparallel}}
\newcommand{\qp}{q_{\shortparallel}}

\newcommand{\kt}{k_{\perp}}
\newcommand{\qt}{q_{\perp}}

\newcommand{\hkp}{\hat{k}_{\shortparallel}}
\newcommand{\hqp}{\hat{q}_{\shortparallel}}

\newcommand{\hkt}{\hat{k}_{\perp}}
\newcommand{\hqt}{\hat{q}_{\perp}}

\newcommand{\tone}{t_{1}}
\newcommand{\ttwo}{t_{2}}

\begin{document}
\title{Photon polarization tensor in a magnetized plasma system}
\author{Jingyi Chao$^{1}$}
\email{jychao@impcas.ac.cn}
\author{Mei Huang$^{2,3}$}
\email{huangm@mail.ihep.ac.cn}
\affiliation{$^{1}$ Institute of Modern Physics, Chinese Academy of Sciences, Lanzhou, 730000, China}
\affiliation{$^{2}$ Institute of High Energy Physics, Chinese Academy of Sciences, Beijing, 100049, China}
\affiliation{$^{3}$ Theoretical Physics Center for Science Facilities, Chinese Academy of Sciences, Beijing, 100049, China}
\date{\today}
%\bigskip
% ======================================================================
\begin{abstract}
We investigate the photon polarization tensor at finite temperature in the presence of a static and homogeneous external magnetic field. In our scheme, the Matsubara frequency summation is performed after Poisson summation, which will be taken easily and convergent quickly in the frame of proper time representation. Moreover, the dependence of Landau levels is expressed explicitly. It demonstrates the convergence of summing Landau levels as it has to be. Consequently, there is no necessary to truncate the Landau level in a numerical estimation. At zero temperature, the Lowest Landau Level (LLL) approximation is analytically satisfied for the imaginary parts of the vacuum photon polarization tensor. Our results examine that, the LLL approximation is not enough for the thermal photon polarization tensor, it gains the contribution not only from the lowest Landau level but also up to the finite-$n$ levels. Such large imaginary ones only show up at finite temperatures, which is the so called Landau damping. It originates from the absorption of soft fields by hard plasma constituents, which is a universal feature of plasma systems. Finally, it was argued that the summation of Matsubara frequency is not commuted with Landau level ones, such conjecture is excluded in our calculations. 
\end{abstract}
% ======================================================================
\maketitle

\section{Introduction}
\label{sec:intro}

At the Relativistic Heavy Ion Collider (RHIC) and the Large Hadron Collider (LHC), strong magnetic fields with the magnitude of $10^{18}\sim 10^{20}~\textmd{G}$ [corresponding to $eB\sim (0.1-1.0\,{\mathrm GeV})^2$], can be generated through non-central heavy-ion collisions~\cite{Skokov:2009qp,Deng:2012pc}.  In recent years, studies of Quantum Chromodynamics (QCD) phase structure and phase diagram under strong external magnetic fields attracted lots of interests \cite{Andersen:2014xxa,Miransky:2015ava,Huang:2015oca}, and it is found that QCD matter under strong magnetic fields exhibits some novel properties, e.g. the chiral magnetic effect (CME) \cite{Kharzeev:2007tn,Kharzeev:2007jp,Fukushima:2008xe}, the magnetic catalysis (MC) \cite{Klevansky:1989vi,Klimenko:1990rh,Gusynin:1995nb} and inverse magnetic catalysis (IMC) \cite{Bali:2011qj, Bali:2012zg, Bali:2013esa}, a possible formation of the vacuum superconductor \cite{Chernodub:2010qx,Chernodub:2011mc} and its survival at high temperature \cite{Liu:2015pna,Liu:2016vuw}, etc. With such strong magnetic fields, Quantum Electrodynamics (QED) will also be modified, such as photon decay into an electron-positron pair via Schwinger mechanism \cite{schwinger1951gauge,tsai1974photon,baier2007pair}, vacuum birefringence of a photon \cite{meszaros1979vacuum,dittrich1998vacuum}, photon splitting and so on \cite{adler1971photon,adler1996photon}.

The strong magnetic field is generated in the early stage of non-central heavy-ion collisions, and then fast decays. It is intriguing to know whether and how properties of the created quark-gluon-plasma (QGP) will be modified by the strong magnetic field, and whether it can be measured in experiment. Dileptons and photons are produced through the bulk of the matter and throughout the entire history of the collision, thus can be regarded as "penetrating probes" of QGP. Recently, experimental evidences of photon anisotropy from the PHENIX Collaboration \cite{Adare:2011zr} challenges existing theoretical models, and it was proposed in \cite{Basar:2012bp} that the photon anisotropy might be induced by a large anisotropic magnetic field in heavy ion collisions. This raises interests of study the photon and dilepton production in the presence of an external magnetic field \cite{Sadooghi:2016jyf,Bandyopadhyay:2016fyd}.

The photon polarization tensor carries the fundamental information of magnetized vacuum or medium \cite{melrose1977polarization,calucci1994nonlogarithmic,hattori2013vacuum,karbstein2013photon,ishikawa2013numerical}. A complete description of the vacuum polarization tensor is particularly complicated to approach, since the vacuum photon polarization tensor is expressed in terms of a double summation of infinite series with respect to two Landau levels occupied by virtual charged particles. Most works were focusing on the strong filed limit with an assumption of Lowest Landau Level (LLL) \cite{calucci1994nonlogarithmic,gusynin1996dimensional}. In \cite{Chao:2014wla} we obtained a full description of vacuum polarization tensor in response to all the Landau levels at any field strength of $B$ for the first time beyond LLL approximation, and we found out that the imaginary part of the photon polarization tensor $\Im\Pi$ becomes nonzero at the time like momenta region $Q^{2}>4(M^{2}+2neB)$ at $T=0$, i.e., the LLL approximation is analytically satisfied \cite{hattori2013vacuum,Chao:2014wla}.

It was not fully understood of the imaginary parts of thermal photon polarization tensor in a magnetized media. The main purpose of this paper is to investigate whether the above conclusion will be influenced by temperatures, which, in turn, is not keeping suitable. At finite temperature, in our scheme, the Matsubara frequency was summed after applying Poisson summation formula, which finished easily and convergent quickly under the help of proper time representation. It was argued that the summation of Matsubara frequency is not commuted with Landau level ones \cite{elizalde1994zeta}. Our calculation exclude this conjecture. The summation of Landau level is convergent as it has to be since no more divergent environment included. However, in the early works, one has to test by a numerical way to find out the cutoff of the Landau level. In our work, the dependence of Landau level is expressed in an obviously matter. Therefore, one is able to truncate the Landau level via a systematic consideration while proceeding numerical simulation.

The paper is organized as follows. We show the vacuum photon polarization tensor $\Pi^{\mu\nu}$ at external $B$ in section \ref{sec:vac}. In section \ref{sec:proper}, we present the magnetized photon polarization tensor $\Pi^{\mu\nu}$ at finite temperature. Due to the additional vector of the velocity of the plasma, the decomposition of the photon polarization tensor will become more complicated. The full expression will be examined in section \ref{sec:PiT}. We discuss the limiting behaviors of thermal photon polarization tensor $\Pi^{\mu\nu}$ and physical explanations at different kinematics regimes in section \ref{sec:thermal}. We end up with the summary and future applications in section \ref{sec:con}.

\section{Photon polarization tensor in the vacuum}
\label{sec:vac}
Following Ref. \cite{gusynin1996dimensional}, the decomposed fermion propagator $\slashed{S}(k)$ in a static and homogeneous external magnetic field can be written as :
\beq
\slashed{S}(k)=i\exp\l(2\hkt^{2}\r)\sum_{n=0}^{\infty}\frac{(-1)^{n}\slashed{D}_{n}(eB,k)}{k_{0}^{2}-k_{3}^{2}-M^{2}-2neB},
\eeq
with
\bea
&&\slashed{D}_{n}(eB,k)=2(\slashed{k}_{\shortparallel}+M)\mathcal{O}^{-}L_{n}(-4\hkt^{2})
\nn
&&-2(\slashed{k}_{\shortparallel}+M)\mc{O}^{+}L_{n-1}(-4\hkt^{2})-4\slashed{k}_{\perp}L_{n-1}^{1}(-4\hkt^{2}).
\eea
Where $\hkt=\kt^{2}/(2eB)$. $L_{n}^{a}(\xi)$ are the generalized Laguerre polynomials, with $L_{n-i}^{a}(\xi)=0$ if $n<i$. $\mc{O}^{\pm}=(1\pm\mi\ga^{1}\ga^{2})/2$ are the projecting operators corresponding to the spin state of charged particle paralleling or anti-paralleling to the direction of external field $B$, and they satisfy following commutation relations:
\begin{eqnarray}
\mc{O}^{\pm}\ga^{\mu}\mc{O}^{\pm}&=&\mc{O}^{\pm}\ga_{\shortparallel} \nonumber \\ \mc{O}^{\pm}\ga^{\mu}\mc{O}^{\mp}&=&\mc{O}^{\pm}\ga_{\perp}.
 \end{eqnarray}
 Here, the metric convention $g^{\mu\nu}$ is decomposed into two orthogonal subspaces
 \begin{equation}
 g^{\mu\nu}_{\shortparallel}=\mathrm{diag}(1,0,0,-1), ~ g^{\mu\nu}_{\perp}=\mathrm{diag}(0,-1,-1,0).
 \end{equation}
 Similar decompositions are adopted for four dimensional momentum $k^{\mu}=k^{\mu}_{\shortparallel}+k^{\mu}_{\perp}$ with
 \begin{equation}
 k^{\mu}_{\shortparallel}=(k^{0},0,0,k^{3}), ~ k^{\mu}_{\perp}=(0,k^{1},k^{2},0)
 \end{equation}
 and Dirac matrices
 \begin{equation}
 \ga^{\mu}_{\shortparallel}=(\ga^{0},0,0,\ga^{3}),~ \ga^{\mu}_{\perp}=(0,\ga^{1},\ga^{2},0).
 \end{equation}

In the vacuum, the photon polarization tensor is expressed as
\begin{equation}
\Pi^{\mu\nu}(q)=-\mi e^2\Tr[\slashed{S}(k)\ga^{\mu}\slashed{S}(p)\ga^{\nu}],
\end{equation}
with $p=k+q$.
Following the procedure described in \cite{Chao:2014wla}, the vacuum polarization tensor can be derived as:
\beq\label{eqn_pi_first}
\Pi^{\mu\nu}=
\int\dif\Ga~I^{\mu\nu}\me^{-\l[\hat{M}^{2}-\eta\hqp^{2}+nx+m(1-x)-\hkp^{2}-i\eps\r]\tau},
\eeq
where $\eta=x(1-x)$ and
\bea
\dif\Ga&=&-\mi e^{2}\sum_{n=0}^{\infty}\sum_{m=0}^{\infty}(-1)^{n+m}\int_{0}^{1}\dif x\int_{0}^{\infty}\dif\tau~\tau
\nn &&
\cdot\int\frac{\dif^{2-\eps}\hkp}{(2\pi)^{2}}\int\frac{\dif^{2}\hkt}{(2\pi)^{2}}
\exp\l[2\hat{k}_{\perp}^{2}+2\hat{p}_{\perp}^{2}\r].
\eea
and the tensor structure of vacuum polarization $I^{\mu\nu}$ becomes
\bea
\label{eqn_Imunu}
I^{\mu\nu}&=&
2\Tr\l[\slashed{k}_{\shortparallel}^{1-x}\ga^{\mu}_{\shortparallel}\slashed{k}_{\shortparallel}^{x}\ga^{\nu}\r]
\l({L}_{n}{L}_{m}+{L}_{n-1}{L}_{m-1}\r)
\nn &&
-2\Tr\l[\slashed{k}_{\shortparallel}^{1-x}\ga^{\mu}_{\perp}\slashed{k}_{\shortparallel}^{x}\ga^{\nu}\r]
\l({L}_{n}{L}_{m-1}+{L}_{n-1}{L}_{m}\r)
\nn &&
-4\Tr\l[\slashed{k}_{\shortparallel}^{1-x}\ga^{\mu}\slashed{k}_{\perp}^{\al}\ga^{\nu}\r]
\l({L}_{n}-{L}_{n-1}\r){L}^{1}_{m-1}
\nn &&
-4\Tr\l[\slashed{k}_{\perp}^{\be}\ga^{\mu}\slashed{k}_{\shortparallel}^{x}\ga^{\nu}\r]
{L}^{1}_{n-1}\l({L}_{m}-{L}_{m-1}\r)\nn &&
+16\Tr\l[\slashed{k}_{\perp}^{\be}\ga^{\mu}\slashed{k}_{\perp}^{\al}\ga^{\nu}\r]
{L}^{1}_{n-1}{L}^{1}_{m-1},
\eea
where
\bea
\slashed{k}_{\shortparallel}^{x}=\slashed{k}_{\shortparallel}+x\slashed{q}_{\shortparallel}+M
&,&
\slashed{k}_{\shortparallel}^{1-x}=\slashed{k}_{\shortparallel}-(1-x)\slashed{q}_{\shortparallel}+M,
\nn
\slashed{k}_{\perp}^{\al}=\slashed{k}_{\perp}+\frac{\al}{\al+\be}\slashed{q}_{\perp}&,&
\slashed{k}_{\perp}^{\be}=\slashed{k}_{\perp}-\frac{\be}{\al+\be}\slashed{q}_{\perp}.
\eea
The augments of $L^{\al,\be}_{n,m}$ have been abbreviated.

\section{Photon polarization tensor at finite temperature}
\label{sec:proper}
% ======================================================================
At finite temperature, the temperature $T$ breaks Lorentz invariance, and we have to separate the zeroth component of the momentum.
To calculate $\Pi^{\mu\nu}$, a general Feynman parameter, $x$, for the denominator factor has been introduced \cite{peskin1995introduction}. Besides, the proper time representation of fermion propagator is employed \cite{schwinger1951green}. Then, the denominator factor is shown as:
\beq
\frac{1}{ab}
=\int_{0}^{1}\dif x\int_{0}^{\infty}\dif\tau~\tau\exp\l[\l(xa+(1-x)b\r)\tau\r]
\eeq
where $a=\kp^{2}-M^{2}-2neB$ and $b=(k+q)_{\shortparallel}^{2}-M^{2}-2meB$ in our work. The longitudinal momentum is denoted as $\kp^2=k_{0}^{2}-k_{3}^{2}$, while the transverse momentum is $\kt^2=-k_{1}^{2}-k_{2}^{2}$. $\tau$ is the variable of proper time. In this paper, we normalize all the energy scale to dimensionless, where $\hat{q}^{2}=q^{2}/(2eB)$, $\hat{M}^{2}=M^{2}/(2eB)$, and so on.

It is well known to shift $k$ to $k-(1-x)q$ can simplify the calculations. But, as pointed out before, because the external magnetic field breaks the uniform of the $3-d$ space, it is only allowed to shift $k_{3}$ to $k_{3}-(1-x)q_{3}$ along the direction of the $B$-field as usual. And the transverse momentum $\kt$ has to be shifted $\kt-\qt\be/(\al+\be)$ shown later by the explicit expression of \Eqn{eqn_summation}. The detail notations $\al,\be$ are described below. We remark that there is no shifting on the zero component since the integration with respect to $k_{0}$ is replaced by summing over Matsubara frequencies $2\mi\pi T\sum_{l}$. Hence, we have
\beq\label{eqn_Pi_munu_one}
\Pi^{\mu\nu}=
\int\dif\Ga~I^{\mu\nu}\exp\l[-\l(\hat{M}^{2}-\eta\hqp^{2}+nx+m(1-x)-\l(\hat{k}_{0}+(1-x)\hat{q}_{0}\r)^{2}+\hat{k}_{3}^{2}\r)\tau\r],
\eeq
where $\eta=x(1-x)$ and the volume space is
\beq
\dif\Ga=-\mi e^{2}\sum_{n=0}^{\infty}\sum_{m=0}^{\infty}(-1)^{n+m}\int_{0}^{1}\dif x\int_{0}^{\infty}\dif\tau\int\frac{\dif^{2}\hkp}{(2\pi)^{2}}\int\frac{\dif^{2}\hkt}{(2\pi)^{2}}\,\tau\exp\l[2\hat{k}_{\perp}^{2}+2\hat{p}_{\perp}^{2}\r].
\eeq
The tensor structure of thermal polarization $I^{\mu\nu}$ is expressed as:
\bea
\label{eqn_I_munu}
I^{\mu\nu}&=&
2\Tr\l[\slashed{k}_{\shortparallel}^{1-x}\ga^{\mu}_{\shortparallel}\slashed{k}_{\shortparallel}^{x}\ga^{\nu}\r]
\l({L}_{n}{L}_{m}+{L}_{n-1}{L}_{m-1}\r)
-2\Tr\l[\slashed{k}_{\shortparallel}^{1-x}\ga^{\mu}_{\perp}\slashed{k}_{\shortparallel}^{x}\ga^{\nu}\r]
\l({L}_{n}{L}_{m-1}+{L}_{n-1}{L}_{m}\r)
\nn &&
-4\Tr\l[\slashed{k}_{\shortparallel}^{1-x}\ga^{\mu}\slashed{k}_{\perp}^{\al}\ga^{\nu}\r]
\l({L}_{n}-{L}_{n-1}\r){L}^{1}_{m-1}
-4\Tr\l[\slashed{k}_{\perp}^{\be}\ga^{\mu}\slashed{k}_{\shortparallel}^{x}\ga^{\nu}\r]
{L}^{1}_{n-1}\l({L}_{m}-{L}_{m-1}\r)
\nn &&
+16\Tr\l[\slashed{k}_{\perp}^{\be}\ga^{\mu}\slashed{k}_{\perp}^{\al}\ga^{\nu}\r]
{L}^{1}_{n-1}{L}^{1}_{m-1}
+2i\Tr\l[\slashed{k}_{\shortparallel}^{1-x}\ga^{1}\ga^{2}\ga^{\mu}_{\perp}\slashed{k}_{\shortparallel}^{x}\ga^{\nu}\r]\l({L}_{n}{L}_{m-1}-{L}_{n-1}{L}_{m}\r)
\nn &&
+4i\Tr\l[\slashed{k}_{\shortparallel}^{1-x}\ga^{1}\ga^{2}\ga^{\mu}\slashed{k}_{\perp}^{\al}\ga^{\nu}\r]
\l({L}_{n}+{L}_{n-1}\r){L}^{1}_{m-1}
+4i\Tr\l[\slashed{k}_{\perp}^{\be}\ga^{\mu}\slashed{k}_{\shortparallel}^{x}\ga^{1}\ga^{2}\ga^{\nu}\r]
{L}^{1}_{n-1}\l({L}_{m}+{L}_{m-1}\r),
\eea
where
\begin{eqnarray}
& & \slashed{k}_{\shortparallel}^{x}=\l(k_{0}+q_{0}\r)\ga^{0}-\l(k_{3}+xq_{3}\r)\ga^{3}+M, \nonumber \\
& & \slashed{k}_{\shortparallel}^{1-x}=k_{0}\ga^{0}-\l(k_{3}+xq_{3}-q_{3}\r)\ga^{3}+M, \nonumber \\
& & \slashed{k}_{\perp}^{\al}=\slashed{k}_{\perp}+\slashed{q}_{\perp}\al/(\al+\be), \nonumber \\
& & \slashed{k}_{\perp}^{\be}=\slashed{k}_{\perp}-\slashed{q}_{\perp}\be/(\al+\be).
 \end{eqnarray}
The augments of $L^{\al,\be}_{n,m}$ have been abbreviated. Under the help of generating function of Laguerre polynomials \cite{gradshteyn2014table}:
\beq
\sum_{n=0}^{\infty}t^{n}L_{n-i}^{a}(\xi)=\frac{t^{i}}{(1-t)^{a+1}}\exp\l[\frac{-t\xi}{1-t}\r]
\eeq
for $|t|<1$, we are able to evaluate the summation of Landau level in a direct manner. We have
\bea
\label{eqn_summation}
&&\exp\l[2\hat{k}_{\perp}^{2}+2\hat{p}_{\perp}^{2}\r]\sum_{n=0}^{\infty}\sum_{m=0}^{\infty}(-1)^{(n+m)}
\exp\l[-(nx+m(1-x))\tau\r]L_{n-i}^{a}(-4\hkt\mb{}^{2})L_{m-j}^{b}(-4\hat{p}_{\perp}^{2})
\nn &&
=\frac{t_{1}^{i}t_{2}^{j}}{(1-\tone)^{a+1}(1-\ttwo)^{b+1}}\exp\l[\frac{\al\be}{\al+\be}\hqt^{2}\r]\exp\l[(\al+\be)\l(\hkt+\frac{\be}{\al+\be}\hqt\r)^{2}\r].
\eea
The last exponential term explains the unusual shifting of transverse momentum which is early taken in \Eqn{eqn_I_munu}. Here, $\tone=-\me^{-x\tau}$, $\ttwo=-\me^{-(1-x)\tau}$, $\al=2(1+\tone)/(1-\tone)$ and $\be=2(1+\ttwo)/(1-\ttwo)$.

% ======================================================================
\section{Tensor Structures at finite temperature}
\label{sec:PiT}
In a thermal plasma, we introduce the velocity of the fluid $u$. It combines with the four momentum of particles to create three second order tensor $u\otimes u, u\otimes q$ and $q\otimes q$. Including the metric $g^{\mu\nu}$, the electromagnetic tensor $F_{\mu\nu}$ and dual tensor $\tilde{F}_{\mu\nu}=\eps_{\al\be\mu\nu}F^{\al\be}$, we have six independent second order tensors. Because of the requirement of Wald identity $\Pi^{\mu\nu}q_{\nu}=0$ and the without loss of generality choice $u^{\mu}u_{\mu}=1$, it means we are able to separate $\Pi^{\mu\nu}$ into four independent structures. From the definition $\Pi^{\mu\nu}=-\mi e^2\Tr[\slashed{S}(k+q/2)\ga^{\mu}\slashed{S}(k-q/2)\ga^{\nu}]=-\mi e^2\Tr[\slashed{S}(k-q/2)\ga^{\nu}\slashed{S}(k+q/2)\ga^{\mu}]$, it draws a conclusion that $\Pi^{\mu\nu}(q)=\Pi^{\nu\mu}(-q)$ \cite{adler1971photon,adler1996photon}. Therefore, $\Pi^{\mu\nu}$ contains symmetry parts made by even powers of four momentum $q$ and antisymmetric parts which formed by odd powers of $q$. To find out the subspace, we set up four mutual orthogonal four momentums:
\begin{equation}
	x_{0}=q^{\mu};\quad x_{1}=\tilde{F}^{\mu\rho}q_{\rho};\quad x_{2}=F^{\mu\rho}q_{\rho};\quad x_{3}=u^{\mu}-x_{0}^{\mu}\frac{u\cdot x_{0}}{x_{0}^2}-x_{1}^{\mu}\frac{u\cdot x_{1}}{x_{1}^2}-x_{2}^{\mu}\frac{u\cdot x_{2}}{x_{2}^2},
\end{equation}
where the fluid velocity $u=(1,0,0,0)$. Hence, the associated transversed symmetric tensors are
\begin{equation}
P_{1}^{\mu\nu}=\frac{x_{1}^{\mu}x_{1}^{\nu}}{x_{1}^{2}};\quad P_{2}^{\mu\nu}=\frac{x_{2}^{\mu}x_{2}^{\nu}}{x_{2}^{2}};\quad P_{3}^{\mu\nu}=\frac{x_{3}^{\mu}x_{3}^{\nu}}{x_{3}^{2}},
\end{equation}
which satisfy the following relationship:
\begin{equation}
P_{i}^{\mu\nu}=P_{i}^{\nu\mu};\quad P_{i}^{\mu\nu}q_{\nu}=0; \quad P_{i}^{2}=P_{i};\quad P_{i}P_{j}=0;\quad \sum_{i=1}^{3}P_{i}^{\mu\nu}=g^{\mu\nu}-\frac{q^{\mu}q^{\nu}}{q^2},
\end{equation}
with $i=1,2,3$ and $i\neq j$. Under the help of the fluid velocity $u$, the antisymmetric tensor must contain the form structures of $u^{\mu}x_{i}^{\nu}-x_{i}^{\nu}u^{\mu}$ for $i=0,1,2$; the natural candidates $F^{\mu\nu}$ and $\tilde{F}^{\mu\nu}$. In fact, we obtain two independent structure forms, which satisfy Wald identity as required. But, one is totally null in our simple electromagnetic tensor environment. Another one is $P_{A}^{\mu\nu}(q)=P_{A}^{\nu\mu}(-q)=u^{\mu}x_{2}^{\nu}-x_{2}^{\mu}u^{\nu}+(u\cdot q)F^{\mu\nu}$. We normalize the conjugate dimensionless projector
\begin{equation}
	P_{4}^{\mu\nu}=\l(P^{\mu\nu}_{4}\r)^{\dagger}=\frac{\mi u^{\mu}x_{2}^{\nu}-\mi x_{2}^{\mu}u^{\nu}}{u\cdot q}+\mi F^{\mu\nu}.
\end{equation}
which satisfy $P_{4}P_{i}=0$ for $i=1,2,3$.

Then, after summing over Landau levels, $\Pi^{\mu\nu}$ takes the form of
\beq
\label{eqn_Pi_munu_two}
\Pi^{\mu\nu}=\sum_{i=1}^{4}P_{i}^{\mu\nu}\pi_{i}=\sum_{i=1}^{4}P_{i}^{\mu\nu}\int\dif\Ga_{E}~\frac{4\tau \me^{-v\tau}}{(1-\tone)(1-\ttwo)}\exp\l[h(x,\tau)\hqt^{2}\r]\mc{I}_{i},
\eeq
where $v=\hat{M}^{2}-\eta\hqp^{2}$ and $h(x,\tau)=\al\be/(\al+\be)$. The scalar functions $\pi_{i}$ are developed by the inner product of $\<P^{\mu\nu}_{i}\colon\Pi^{\mu\nu}\>$, which are in the form of as below
\bea
\mc{I}_{1}&=&2\l(M^{2}+\Th_{1}^{2}\r)(1+\tone\ttwo)
	+\frac{4\l[\al\be\qt^{2}-\kt^{2}(\al+\be)^{2}\r]\tone\ttwo}{(\al+\be)(1-\tone\ttwo)};
	\nonumber \\
\mc{I}_{2}&=&2\Th_{2}^{2}(\tone+\ttwo)+\frac{4\qt^{2}\al\be\tone\ttwo}{(\al+\be)(1-\tone\ttwo)};
\nn
\mc{I}_{3}&=&\frac{2\qt^{2}}{q^{2}}\l(M^{2}-\Th_{1}^{2}\r)(1+\tone\ttwo)
    +\frac{4\qt^{2}\l[\al\be(\qt^{2}-\qp^{2})-\kt^{2}(\al+\be)^{2}\r]\tone\ttwo}{q^{2}(\al+\be)(1-\tone\ttwo)}
\nn &&
	+\frac{4\qt^{2}}{q^{2}(1-\tone\ttwo)}\l[\l(\tilde{k}_{0}q_{0}-(1-x)\qp^{2}\r)(1-\tone^{2})\ttwo-(\tilde{k}_{0}q_{0}+x\qp^{2})(1-\ttwo^{2})\tone\r]+\frac{2\qp^{2}\Th_{2}^{2}}{q^{2}}(\tone+\ttwo);
\nn
\mc{I}_{4}&=&4\Th_{2}^{2}(\tone-\ttwo)-\frac{4\qt^{2}}{q_{0}(1-\tone\ttwo)}\l[\l(\tilde{k}_{0}-(1-x)q_{0}\r)(1+\tone)^{2}\ttwo+\l(\tilde{k}_{0}+xq_{0}\r)(1+\ttwo)^{2}\tone\r];	
\eea
where $\Th_{1}^{2}=\eta\qp^{2}-\l(q_{0}^{2}+q_{3}^{2}\r)\l(\tilde{k}_{0}^{2}+k_{3}^{2}\r)/\qp^{2}+(1-2x)\tilde{k}_{0}q_{0}$, $\Th_{2}^{2}=\tilde{k}_{0}^{2}-(1-2x)\tilde{k}_{0}q_{0}-k_{3}^{2}-M^{2}-\eta\qp^{2}$ and $\tilde{k}_{0}=k_{0}+(1-x)q_{0}$. The identity $(1-\tone)(1-\ttwo)(\al+\be)=4(1-\tone\ttwo)$ has been plugged in. After performing wick rotation, one has $k^2=-k_{E}^{2}$ and the integral volume space is
\beq
\dif\Ga_{E}=e^{2}\int_{0}^{1}\dif x\int_{0}^{\infty}\dif\tau\int\frac{\dif\Ga_{\kp}}{(2\pi)^{2}}\int\frac{\dif\Ga_{\kt}}{(2\pi)^{2}},
\eeq
where
\bea
\int\dif\Ga_{\kp}&=&\hat{T}\sum_{l=-\infty}^{\infty}\exp\l[-\l(\hat{\om}_{l}-i(1-x)\hat{q}_{0}\r)^{2}\tau\r]\int\dif\hat{k}_{3}\exp\l[-\hat{k}_{3}^{2}\tau\r],
\nn
\int\dif\Ga_{\kt}&=&\int\dif^{2}\hkt\exp\l[-(\al+\be)\l(\hat{k}_{1}^{2}+\hat{k}_{2}^{2}\r)\r].
\eea
Note here $k_{0}$ has been replaced by $\mi\om_{l}$, where $\hat{\om}_{l}^{2}=(2l+1)^{2}\pi^{2}T^{2}/(2eB)$.

With respect to the momenta $\kt^{2}$, the integrand of $\Pi^{\mu\nu}$ is in a simple Gaussian form. Hence, we are able to write down the polarization tensor by some simple scalar functions in Euclidean space. To simplified, four scalar functions $\Xi$ are denoted as below
\beq\label{eqn_Xi}
\Xi(\vartheta,\varrho|v)=\frac{\hat{k}_{3}^{2\vartheta}}{16\pi^{3}}\sum_{l=-\infty}^{\infty}\int_{0}^{\infty}\dif\tau\frac{\hat{T}\tau\l(\mi \hat{T}\r)^{\varrho}(l+\varpi)^{\varrho}}{1-\me^{-\tau}}\exp\l[-\l(v+\hat{k}_{3}^{2}+(l+\varpi)^{2}\hat{T}^{2}\r)\tau+h(x,\tau)\hqt^2\r].
\eeq
Here $\hat{T}=(2\pi T)/\sqrt{2eB}$ and $\varpi=\frac{1}{2}-\mi(1-x)q_{0}/(2\pi T)$. And then we let $\Xi_{i}(v)=\Xi(\vartheta=0,\varrho=i;v)$ for $i=0,1,2$ and $\Xi_{3}(v)=\Xi(\vartheta=1,\varrho=0;v)$. Also, we have some scalar functions are in the form as
\begin{equation}\label{eqn_varXi}
	\varXi_{\varrho}(v)=\frac{1}{16\pi^{3}}\sum_{l=-\infty}^{\infty}\int_{0}^{\infty}\dif\tau\frac{\hat{T}\tau\l(\mi \hat{T}\r)^{\varrho}(l+\varpi)^{\varrho}}{\l(1-\me^{-\tau}\r)^{2}}\exp\l[-\l(v+1+\hat{k}_{3}^{2}+(l+\varpi)^{2}\hat{T}^{2}\r)\tau+h(x,\tau)\hqt^2\r]
\end{equation}
for $\varrho=0,1$. Step on, Poisson summation formula have been applied in \Eqn{eqn_Xi} and \Eqn{eqn_varXi}, such as
\begin{equation}\label{eqn_poisson}
	\sum_{l=-\infty}^{\infty}\exp\l[-\l(l+\varpi\r)^{2}\frac{\pi}{\si}\r]=\sum_{l=-\infty}^{\infty}\si^{\frac{1}{2}}\exp\l[-l^{2}\pi\si-2\mi l\pi\varpi\r],
\end{equation}
%\bea
%	\sum_{l=-\infty}^{\infty}\exp\l[-\l(l+\varpi\r)^{2}\frac{\pi}{\si}\r]&=&\sum_{l=-\infty}^{\infty}\si^{\frac{1}{2}}\exp\l[-l^{2}\pi\si-2\mi l\pi\varpi\r].
%	\nn
%	\sum_{l=-\infty}^{\infty}(l+\varpi)\exp\l[-\l(l+\varpi\r)^{2}\frac{\pi}{\si}\r]&=&\sum_{l=-\infty}^{\infty}\mi l\si^{\frac{3}{2}}\exp\l[-l^{2}\pi\si-2\mi l\pi\varpi\r],
%	\nn
%	\sum_{l=-\infty}^{\infty}(l+\varpi)^{2}\exp\l[-\l(l+\varpi\r)^{2}\frac{\pi}{\si}\r]&=&\frac{1}{2\pi}\sum_{l=-\infty}^{\infty}\l(\si^{\frac{3}{2}}-2\pi l^{2}\si^{\frac{5}{2}}\r)\exp\l[-l^{2}\pi\si-2\mi l\pi\varpi\r],
%\eea
and so on. In this work, we have $\si=\pi/(\tau\hat{T}^{2})$.

Obviously, the summation over $l$ from $-\infty$ into $\infty$ is able to split into positive values and zero, i.e. the thermal, $\Xi_{i}^{T}$ and vacuum parts, $\Xi_{i}^{0}$, respectively. All the results of vacuum contribution have been discussed in \cite{Chao:2014wla}. In this letter, we focus on the thermal parts.
% ======================================================================
\section{Thermal Results for the photon polarization tensor}
\label{sec:thermal}

The thermal scalar functions are expressed as below
\begin{equation}
	\label{eqn_Xi_thermal}
	\Xi_{i}^{T}(v)=\frac{\hat{k}_{3}^{2\vartheta}}{16\pi^{\frac{5}{2}}}\sum_{s=\pm 1}\sum_{l=1}^{\infty}\int_{0}^{\infty}\dif\tau\frac{\tau^{\frac{1}{2}}s^{\varrho}G(\varrho)}{2\tau(1-\me^{-\tau})}\exp\l[-\l(v+\hat{k}_{3}^{2}\r)\tau-\frac{\pi^{2}l^{2}}{\hat{T}^{2}\tau}+h(x,\tau)\hqt^2-sl\xi\r]\cos(l\pi),
\end{equation}
where $\xi=(1-x)q_{0}/T$. We note here the explicit forms of $G(\varrho)=2\tau(-\pi l/\hat{T}\tau)^{\varrho}-\Ga(\varrho+1)+1$. We rephrase the above equation via the geometirc trick $1/(1-\me^{-\tau})=\sum_{n=0}\me^{-n\tau}$. Besides, the thermal parts of $\varXi_{i}^{T}$ are easily addressed by $\me^{-\tau}/(1-\me^{-\tau})^{2}=\sum_{n=1}n\me^{-n\tau}$ for $i=0,1$.

Now, the integral with respect to $\tau$ can be written in terms of a modified Bessel function of second kind, $K_{\nu}(y)$, where
\begin{equation}\label{eqn_bessel_K}
	\int_{0}^{\infty}\tau^{\nu-1}\exp\l(-\la\tau-\frac{\chi}{\tau}\r)\dif\tau=2\l(\frac{\chi}{\la}\r)^{\frac{\nu}{2}}K_{\nu}\l(2\sqrt{\la\chi}\r),
\end{equation}
for $\Re\la >0$ and $\Re\chi >0$. In strong $B$-field, the most contribution of the integration with respect to $\tau$ is from the regime where $\tau\sim (\chi/\la)^{\frac{1}{2}}\sim\sqrt{2eB}$. It means we are able to set $h(x,\tau)=1$ safely. The details reason can be found in the reference \cite{Chao:2014wla}.

The special case, $\nu=\frac{1}{2}$, of the modified Bessel function at the second find is $K_{\frac{1}{2}}(\mr{z})=\sqrt{\frac{\pi}{2\mr{z}}}\me^{\mr{-z}}$. We apply the saddle point method to integrate with respect to $x$,
\begin{equation}\label{eqn_int_x}
 	\sum_{l=1}^{\infty}\sum_{s=\pm 1}\int_{0}^{1}l^{\frac{1}{2}}y^{-\frac{1}{4}}K_{\frac{1}{2}}(l\sqrt{y})\exp\l(sl\xi\r)\cos(l\pi)\dif x\simeq \sum_{s=\pm 1}\frac{\me^{\frac{\mi(\pi-\phi)}{2}}\mr{Li}_{0}\l(-\me^{f(x_{0})}\r)}{\sqrt{y(x_{0})}}\l(\frac{2\pi }{\lvert f^{(2)}(x_{0})\rvert}\r)^{\frac{1}{2}}
 \end{equation}
where $f(x)=-\sqrt{y}+s\xi$, $y(x)=\l(E_{k}^{2}-\eta\qp^{2}\r)/T^2$ and $E_{k}^{2}=M^2+2neB+k_{3}^{2}$. $x_{0}$ is the saddle point where $f^{(1)}(x_{0})=0$. $\phi$ is the angle determined by which the contour passes through the saddle point. When $f^{(2)}(x_{0})=\lvert r\rvert\pm\mi\ep$, $\phi$ is equal to $0$ or $2\pi$ for positive or negative sign, respectively. To obtain a real root of $x_{0}$ locating in the interval $(0,1)$, it requires $\qp^{2}\leq 4E_{k}^{2}\leq \qp^{4}/q_{0}^{2}$. Therefore, there is no saddle point in the time like momenta for $\qp^{2}>0$. Indeed, at the energy region $\qp^{2}>4E_{k}^{2}$, \Eqn{eqn_bessel_K} is not applicable because $\la$ will be not positive. We will consider this complexity issue in the next later. Here, we only include the result for space like momenta. And then it means $\qp^2$ being less than $-2\sqrt{M_{n}^{2}q_{0}^{2}}$, where $M_{n}^{2}=M^2+2neB$. Clearly, for $\qp^{2}<0$, the saddle point locates at
\begin{equation}
	x_{0}=\frac{1}{2}-s\frac{\sign(q_{0})}{2}\sqrt{\frac{4E_{k}^{2}q_{0}^{2}-\qp^{2}q_{0}^{2}}{\qp^{4}-\qp^{2}q_{0}^{2}}}.
\end{equation}
It leads $Tf^{(2)}(x_{0};q_{0}\to q_{0}\pm\mi\ep)=2\lvert q_{3}\rvert^{3}/\sqrt{\qp^4-4E_{k}^{2}\qp^{2}}\pm\mi\sign(q_{0})\ep$ being always positive for $s=\pm 1$. And then the answer of \Eqn{eqn_int_x} becomes pure imaginary. Using the notation of $\<f \>=\int f\dif x\dif\hat{k}_{3}$, one has
\bea\label{eqn_Xi_T1}
	\l\<\eta\qp^{2}\Xi_{0}^{T}(v\pm\mi\ep)\r\>&=&\frac{\pm\mi}{8\pi^{\frac{3}{2}}}\sum_{s=\pm 1}\sum_{n=0}\int_{M_{n}^{2}}^{\frac{\qp^{4}}{4q_{0}^{2}}}\frac{T^{\frac{1}{2}}(2eB)\qp^{2}}{\lvert q_{3}\rvert^{\frac{1}{2}}\l(\La-M_{n}^{2}\r)^{\frac{1}{2}}}
	\frac{\p}{\p \qp^{2}}\frac{\mr{Li}_{0}\l[-\exp\l(\frac{s q_{0}}{2T}-\frac{\sqrt{\qp^{4} q_{3}^{2}-4\La \qp^{2}q_{3}^{2}}}{2\qp^{2} T}\r)\r]}{\l(\qp^4-4\La \qp^{2}\r)^{\frac{1}{4}}}
	\dif \La
	\nn
	&\simeq&\frac{\mp\mi}{8\pi^{\frac{3}{2}}}\sum_{n=0}^{j}\frac{(2eB)\l(\qp^{4}-4q_{0}^{2}M_{n}^{2}\r)^{\frac{1}{2}}}{3T^{\frac{1}{2}}\lvert q_{0}\rvert^{\frac{3}{2}}}\mr{Li}_{-1}\l(-\me^{-\frac{q_{3}^{2}}{2\lvert q_{0}\rvert T}}\r)
\eea
where $j=\lfloor \qp^4/(8eBq_{0}^{2})-\hat{M}^{2}\rfloor$.

In this paper, the requirement $\Re\chi >0$ of \Eqn{eqn_bessel_K} is satisfied always. $\la$ is non-positive when $\qp^2>4M_{n}^2$, which induces instability of exponential integration. Considering such case, we express the result by the Cauchy's principal value method,
\begin{equation}
	\Im\l[\int_{0}^{\infty}\tau^{\nu-1}\exp\l(\la\tau-\frac{\chi}{\tau}\r)\dif\tau\r]=\pi\l(\frac{\chi}{\la}\r)^{\frac{\nu}{2}}J_{\nu}\l(2\sqrt{\la\chi}\r),
\end{equation}
where $J_{\nu}(\mr{z})$ is the Bessel functions of the first kind; $\Re\la >0$ and $\Re \chi >0$. Due to the formula of $J_{\nu}\l(\me^{\mi m\pi}z\r)=\me^{\mi m\nu\pi}J_{\nu}(z)$ for interger $m$, one gets $J_{\nu}(z)|_{p_{0}\to p_{0}+\mi\ep}=-J_{\nu}(z)|_{p_{0}\to p_{0}-\mi\ep}$ for half interger $\nu$.

At $\nu=\frac{1}{2}$, one has $J_{\frac{1}{2}}(\mr{z})=\sqrt{\frac{2}{\pi\mr{z}}}\sin\mr{z}$. Therefore,  we obtain
\begin{equation}
\label{eqn_sum_BesselJ}
	\sum_{s=\pm 1}\sum_{l=1}^{\infty}l^{\frac{1}{2}}\lvert y\rvert^{-\frac{1}{4}}J_{\frac{1}{2}}\l(l\sqrt{-y}\r)\exp\l(sl\xi\r)\cos(l\pi)
		=\sqrt{\frac{2}{\pi}}\frac{-2\sin\sqrt{-y}}{\sqrt{-y}\l(\me^{\xi}+2\cos\sqrt{-y}+\me^{-\xi}\r)},
\end{equation}
which is exponentially suppressed in the low temperature limit for $\lvert q_{0}\rvert\gg T$. Hence, we will render only the results at high temperatures. Obviously, in the strong magnetic fields, the allowed momenta region $q_{0}^{2}>\qp^{2}>2neB\gg T^{2}$ is extremely precluded at finite $n$. The lowest Landau level is left and contributes an imaginary part in the time like momenta, which is similar to the results at zero temperature. Meanwhile, for $\xi\ll 1$ and $\sqrt{-y}\ll 1$, one has $\me^{\xi}+2\cos\sqrt{-y}+\me^{-\xi}\simeq 4$ and gets
\begin{equation}
	\l\<\l(\eta\qp^{2}-M^{2}\r)\Xi_{0}^{T}(v\pm\mi\ep)\r\>
	\simeq\mp\frac{\mi(2eB)}{8\pi^{2}T}\int_{\frac{1-\bar{x}}{2}}^{\frac{1+\bar{x}}{2}}\dif x\int_{0}^{\bar{k}_{3}}\l(\frac{\qp^{2}\p}{\p \qp^{2}}+\frac{M^{2}\p}{\p M^{2}}\r)\frac{\sin\sqrt{-y}}{\sqrt{-y}}\dif k_{3}
	=\pm\frac{\mi(2eB)\l(\qp^{2}-4M^{2}\r)^{2}}{2^{11}\cdot\pi T^{3}\lvert \qp\rvert},
\end{equation}
where $\bar{x}=\sqrt{1-4M^{2}/\qp^2}$ and $\bar{k}_{3}=\sqrt{\eta\qp^{2}- M^{2}}$.

It is easy to see we have two branch cuts along the real axis of $q_{0}$. After extended it to the complex plane, one has $\mr{Disc}\,\pi_{i}(q_{0})=\pi_{i}(q_{0}+\mi\ep)-\pi_{i}(q_{0}-\mi\ep)=2\mi\,\Im \pi_{i}(q_{0})$. In the strong magnetic field and chiral limit, 
we write down the discontinuities directly.

The first branch cut, $q_{0}^{2}<q_{3}^{2}+2M_{n}^{2}-2\sqrt{M_{n}^{4}+q_{3}^{2}M_{n}^{2}}<q_{3}^{2}$, is developed at finite temperatures, which is corresponding to the process of $\ga+\psi\leftrightharpoons\psi$. One has:
\bea\label{eqn_space_pi}
	\mr{Disc}\,\pi_{1}(q_{0})&\simeq&2\,\Im \l\<\l(M^{2}+\eta\qp^{2}\r)\l(\Xi_{0}^{T}(v)+\Xi_{0}^{T}(v+1)\r)\r.
	\nn &&
	\l.+(2eB)\l(\Xi_{2}^{T}(v)+\Xi_{2}^{T}(v+1)+\Xi_{4}^{T}(v)+\Xi_{4}^{T}(v+1)+2\varXi_{0}^{T}(v)\r)\r\> \nonumber \\
    &\simeq&\l(\sum_{n=0}^{j}+\sum_{n=1}^{j}\r)\frac{-1}{4\pi^{\frac{3}{2}}}\frac{(2eB)\l(\qp^{4}-4q_{0}^{2}M_{n}^{2}\r)^{\frac{1}{2}}}{T^{\frac{1}{2}}\lvert q_{0}\rvert^{\frac{3}{2}}}\mr{Li}_{-1}\l(-\me^{-\frac{q_{3}^{2}}{2\lvert q_{0}\rvert T}}\r)
    \nn
	\mr{Disc}\,\pi_{2}(q_{0})&\simeq&\frac{\qt^{2}}{2eB}\mr{Disc}\,\pi_{1}(q_{0});\nonumber \\
	\mr{Disc}\,\pi_{3}(q_{0})&\simeq&-\frac{\qt^{2}}{3q^{2}}\mr{Disc}\,\pi_{1}(q_{0});\nonumber \\
	\mr{Disc}\,\pi_{4}(q_{0})&=&0.
\eea
Where $\mr{Disc}\,\Pi^{\mu\nu}(q_{0})=\sum_{i=1}^{4}P_{i}^{\mu\nu}\,\mr{Disc}\,\pi_{i}(q_{0})$ as denoted before. Roughly speaking, the finite-$n$ Landau level contribution is supposed to exponentially suppressed, $\me^{-n}$, which is the underlying of the approximation of LLL. But, in a strictly manner, the $n$-th Landau levels present as $L_{n}^{(\al)}\me^{-n}$. Indeed, the Laguerre polynomial was neglected improperly in lots of early works. For large $n$, the asymptotic behavior of $L_{n}^{(\al)}$ is limit to $n^{\frac{\al}{2}-\frac{1}{4}}$. Hence, $L_{n}^{(\al)}\me^{-n}$ is characterized by a non-monotonic behavior of $n$ when $\al\geq 1$. In other words, the LLL approximation will break down if the terms contained $L_{n}^{(1)}$ play an important role in the estimation.

The second branch cut, $q_{0}^{2}>q_{3}^{2}+4M^{2}$, is due to the conventional process $\ga\leftrightharpoons\psi+\psi$. In the chiral limit $M\to 0$, One has:
\bea\label{eqn_time_pi}
	\mr{Disc}\,\pi_{1}(q_{0})&\simeq&2\,\Im
	\l\<\l(\eta\qp^{2}\r)\Xi_{0}^{T}(v)-(2eB)\l(\Xi_{2}^{T}(v)+\Xi_{4}^{T}(v)\r)\r\>
    =\frac{(2eB)\lvert \qp\rvert}{2^{11}\cdot\pi T^{3}};
	\nn
	\mr{Disc}\,\pi_{2}(q_{0})&=&0;\nonumber \\
	\mr{Disc}\,\pi_{3}(q_{0})&=&-\frac{\qt^{2}}{q^{2}}\mr{Disc}\,\pi_{1}(q_{0});\nonumber \\
	\mr{Disc}\,\pi_{4}(q_{0})&=&0.
\eea

\section{Conclusion and Discussions}
\label{sec:con}

In this work, we have completed the form of the photon polarization tensor at finite temperature in the media of a static and homogeneous magnetic field. In our process, the summing of Matsubara frequency was applied elegantly, since its concise formulation limit well and achieve easily in the approach of proper time representation. Meanwhile, it was argued that the summation of Matsubara frequency is not commuted with Landau level ones, such problem does not occur in our calculations. As expected, without a divergent environment included, the summation of Landau level is convergent. Before, one has to attend the cutoff of the Landau level via a test of numerical method. Unlike such cumbersome way, we truncate the Landau level in a systematic consideration since the dependence of Landau level is expressed in an obviously manner.

It is well known that the self-energies of gauge bosons take the same forms for both QED and QCD plasmas in the limit of long wavelength~\cite{PhysRevD.91.025013}. It takes place in the strong magnetic filed limit, as well. In the space like momenta regime $q_{3}^{2}\gg q_{0}^{2}$ and strong $B$-fields limit $(2eB)\gg q_{0}^{2},\,q_{3}^{2}$, it is allowed us to define $(2eB)\sim\la T^{2}$ and $\la>1$. The classification of the energy scale is similar to hard-loop action, where loop momenta $k\sim M_{n}$ for finite-$n$ Landau levels while as the external momenta $q_{3}\sim\la^{-\frac{1}{2}}T^{\frac{1}{2}}M_{n}^{\frac{1}{2}},~q_{0}\sim\la^{-\frac{3}{2}}T$. Described by the result of \Eqn{eqn_space_pi}, $\mr{Disc}\,\pi_{1}$ is at the order of $\la^{\frac{7}{4}}\mr{Li}_{-1}(-\me^{-\la})$, which is not monotonically decreasing as $\la$ increasing. Physically, such unique feature are essentially same as other gauge theories governed plasma systems. This large imaginary part only arises at finite temperatures, which is the so called Landau damping. It demonstrates the absorption of soft fields by hard plasma constituents. Such kind of processes gains the contribution not only from the lowest (soft) Landau level but also up to the finite-$n$ (hard) levels. The general mathematic explanations have been analyzed as before. Eventually, we conclude that the LLL approximation is suit at zero temperature but not well described at the finite temperatures. In particular, the hard-loop approximation takes control for the magnetized plasma systems, whose constituent of typical momentum is much larger than the probing wave vector.

Our results have two-fold applications in the next future. One is that the Landau damping effect from the finite Landau levels has the potential to solve the inverse magnetic catalysis which observed in the lattice simulation. The second is that the modification of the photon production rate is a new effect which has to be considered for discussing photon $v_{2}$ puzzle in heavy ion collisions.

% ======================================================================
\vspace{1.5em}
{\bf\it Acknowledgement.---}
We thank I.Shovkovy for discussion. This work is founded by MOST 2015CB856903, supported by National Key Basic Research Program (NKBRP), the NSFC under Grant No. 11621131001, DFG and NSFC (CRC 110), CAS key project KJCX2-EW-N01.

%\begin{thebibliography}

\end{document}